\documentclass[namedreferences]{solarphysics}

\usepackage[optionalrh]{spr-sola-addons} % For Solar Physics 
\usepackage{graphicx}        % For eps figures, newer & more powerfull
\usepackage{amssymb}        % useful mathematical symbols
\usepackage{color}           % For color text: \color command
\usepackage{breakurl}        % For breaking URLs easily trough lines
\usepackage[colorlinks=true,citecolor=blue]{hyperref}
\usepackage{epstopdf}

\newcommand{\arcsec}{$^{''}$}

\newcommand{\aap}{    {\it Astron. Astrophys.}}

\newcommand{\aj}{     {\it Astron. J.}} 
\newcommand{\apj}{    {\it Astrophys. J.}}
\newcommand{\apjl}{   {\it Astrophys. J. Lett.}}
\newcommand{\apjs}{   {\it Astrophys. J. Suppl.}}

\newcommand{\solphys}{{\it Solar Phys.}}
 
\newcommand{\ssr}{    {\it Space Sci. Rev.}} 
\chardef\us=`\_

\begin{document}

\begin{article}
\begin{opening}

\title{Simultaneous longitudinal and transverse oscillation in an active filament}

\author[addressref={aff1},corref,email={vaibhav@iiap.res.in}]{\inits{Vaibhav}\fnm{Vaibhav}~\lnm{Pant}}%\sep
\author[corref,addressref={aff1,aff2},email={}]{\inits{Rakesh}\fnm{Rakesh}~\lnm{Mazumder}}%\sep

\author[addressref={aff3},email={}]{\inits{Yuan}\fnm{Ding}~\lnm{Yuan}}%\sep
\author[addressref={aff1,aff2},email={}]{\inits{Dipankar}\fnm{Dipankar}~\lnm{Banerjee}}%\sep
\author[addressref={aff4},email={}]{\inits{Abhishek}\fnm{Abhishek K.}~\lnm{Srivastava}}%\sep
\author[addressref={aff5},email={}]{\inits{Yuandeng}\fnm{Yuandeng}~\lnm{Shen}}%\sep

\address[id=aff1]{Indian Institute of Astrophysics, Koramangala, Bangalore 560034.}
\address[id=aff2]{Center of Excellence in Space Sciences, IISER Kolkata, Mohanpur 741246, West Bengal, India}
\address[id=aff3]{Jeremiah Horrocks Institute, University of Central Lancashire, UK}
\address[id=aff4]{Indian Institute of Technology (BHU), Varanasi, India}
\address[id=aff5]{Yunnan Observatories, Chinese Academy of Sciences, Kunming 650216, China}
%\address[id=aff3]{BHU}

\runningauthor{V. Pant et al.}
\runningtitle{Oscillation in filament}

\begin{abstract}
We report on the co-existence of longitudinal and transverse oscillations in an  active filament. On March 15$^{th}$ 2013, a M1.1 class flare was observed in the active region AR 11692. A CME was found to be associated with the flare. {The CME generated a shock wave that triggered the oscillations in a nearby filament}, situated at the south-west of the active region as observed from National Solar Observatory (NSO)\textit{Global Oscillation Network Group}(GONG) H$\alpha$ images. In this work we report the longitudinal oscillations in the two ends of the filament, co-existing with the transverse oscillations. We propose a scenario in which {an} incoming shock wave hits the filament obliquely and triggers both longitudinal and transverse oscillations. Using the observed parameters, we estimate the lower limit of the magnetic field strength. We use simple pendulum model with gravity as the restoring force to estimate the radius of curvature. We also calculate the mass accretion rate which causes the filament motions to damp quite fast.
\end{abstract}
\end{opening}
%-------------------------------------------------

\section{Introduction}
Filaments support both longitudinal and transverse oscillations \citep[see,][]{2012LRSP....9....2A}.  The oscillations in the filaments can be used to diagnose the local plasma conditions and magnetic field by applying the principle of MHD seismology \citep[see,][]{2005LRSP....2....3N, 2005ApJ...624L..57A}. 
These oscillations are broadly classified as large amplitude \citep{2009SSRv..149..283T} and small amplitude \citep{2002SoPh..206...45O} oscillations. Large amplitude transverse oscillations, where a filament oscillates as whole, are often associated with the disturbances coming from nearby flares \citep{1966AJ.....71..197R,2006A&A...449L..17I,2004ApJ...608.1124O,2011A&A...531A..53H,2013ApJ...777..108S,2015RAA....15.1713P}.  In contrast, the large amplitude longitudinal oscillations are always found to be associated with small energetic events like a subflare at one leg of the filament \citep{2003ApJ...584L.103J}, a small flare in a nearby active region \citep{2006SoPh..236...97J,2007A&A...471..295V,2012ApJ...760L..10L}  or jets \citep{2014ApJ...785...79L}. 
There have been several reports on large amplitude longitudinal oscillations \citep{2003ApJ...584L.103J, 2006SoPh..236...97J,2007A&A...471..295V,2012ApJ...750L...1L, 2012ApJ...757...98L, 2012A&A...542A..52Z,2014ApJ...785...79L}. \inlinecite{2012ApJ...750L...1L} developed a 1D self-consistent model which explained both restoring force and damping of large amplitude longitudinal oscillations. The model is analogous to an oscillating pendulum assuming the radius of curvature of magnetic dip as the length of the pendulum. \inlinecite{2014ApJ...785...79L} used this model to explain the strong damping of large amplitude longitudinal oscillations observed in the filament threads. Recently, \inlinecite{2016ApJ...817..157L} performed 2D non-linear time dependent MHD simulations where a magnetic field can get distorted in response to the mass loading. The authors have reported that a simple pendulum model can be used to characterise the large amplitude longitudinal oscillations even in a scenario where the magnetic field responds to the plasma motions.  \\
Damping of oscillations have been observed in large amplitude transverse oscillations \citep{2011A&A...531A..53H,2015RAA....15.1713P}. \inlinecite{2006RSPTA.364..433G,2007A&A...463..333A} used seismology combining the damping time and period of oscillations in a consistent manner to estimate the magnetic field strength and inhomogeneity length scale in the coronal loops.  \inlinecite{2008ApJ...682L.141A} applied the same technique to the oscillating threads of a prominence. \inlinecite{2015RAA....15.1713P} applied this technique to estimate magnetic field and inhomogeneity length scale of a filament as whole. \\
In addition to transverse oscillations, large amplitude longitudinal oscillations are also found to be damped \citep{2012ApJ...750L...1L,2014ApJ...785...79L}. Unlike transverse oscillations, that are now believed to be damped by the resonant absorption, damping mechanisms in longitudinal oscillations are not well understood. Several  damping mechanisms have been proposed, {\it e.g.,} energy leakage \citep{1969SoPh....6...72K}, dissipation \citep{2009SSRv..149..283T}  but these mechanisms are under debate. However, mass accretion due to the condensation of filament material \citep{2012ApJ...750L...1L,2014ApJ...785...79L,2016ApJ...817..157L,2016arXiv160503376R} is shown to be promising in explaining the damping of longitudinal oscillations in a filament. Recently, \inlinecite{2016arXiv160503376R} have reported the evidence of strong damping of longitudinal oscillations in the filament threads that are modelled as a curved magnetic tube.  \\ %Filament eruption often produces CME, hence important for space weather prediction. It has been found that large scale oscillation is often a precursor of filament eruption. Hence study of large scale filament oscillation is extremely important.\\
\inlinecite{2014ApJ...795..130S} reported that an incoming shock wave can trigger longitudinal and transverse oscillations in the filaments depending on the interaction angle between the filament and the shock wave. \inlinecite{2008ApJ...685..629G} has reported coexistence of two transverse oscillation modes, one in the plane of sky and other along the line of sight, in a filament in response to a Moreton wave. Both modes were perpendicular to the filament axis. In this work, we report a unique observation of the co-existence of large amplitude damped longitudinal oscillations and large amplitude damped transverse oscillations in an active region filament. 
The paper is organised as follows. In Section 2, we describe the observations. In Section~3 we discuss the methods of data analysis. We present the results in Section 4, which is followed by the discussion and conclusions in Section 5.

\section{Observation}

\begin{figure}[h]
\centering
\includegraphics[scale=0.81,angle=90]{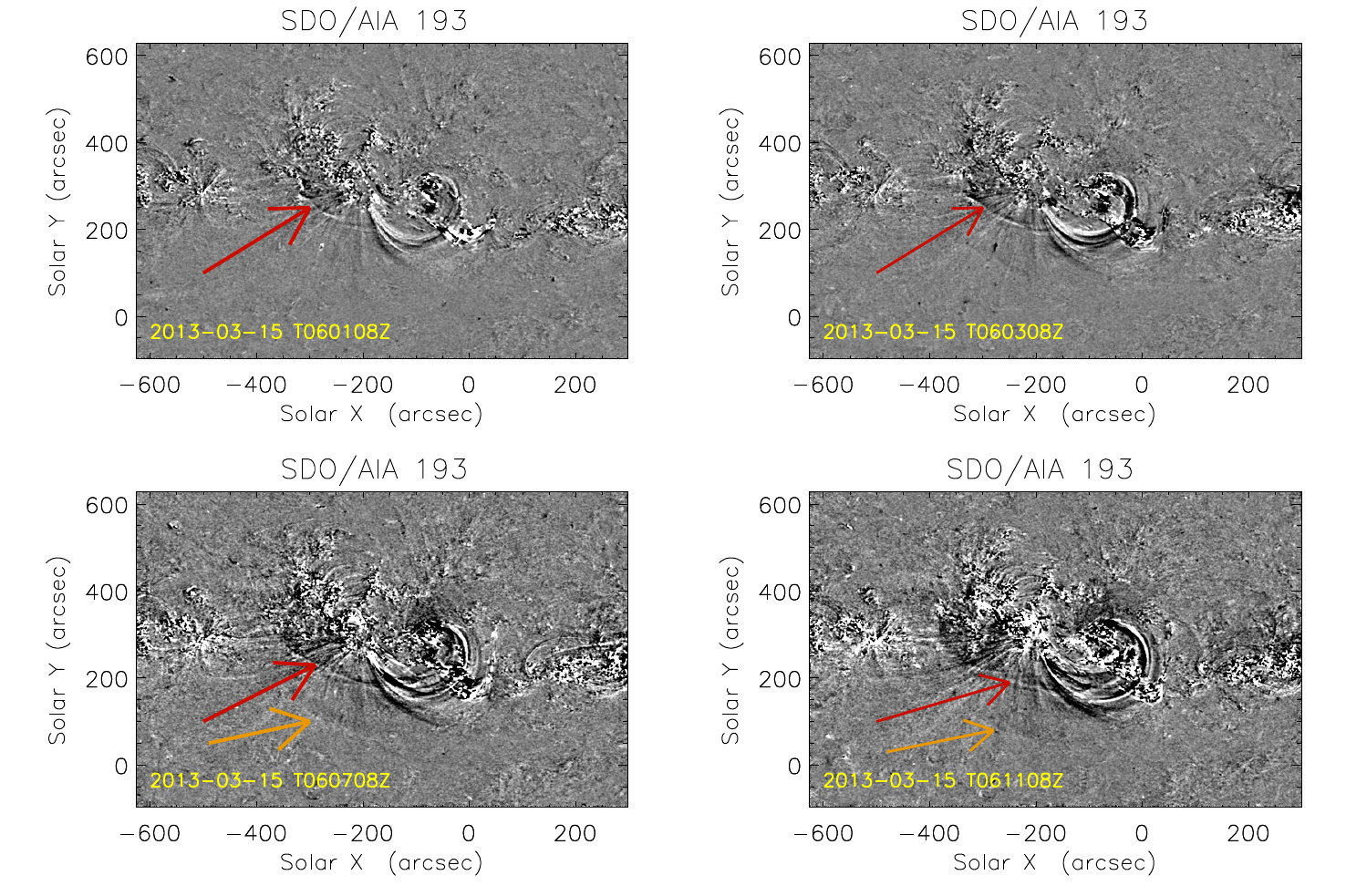}
\caption{Difference images of AIA~193~\AA~at four different time instances. Arrows in red and green represent the shock front.}
\label{shock_fig}
\end{figure}

A M1.1 Class flare was observed by GOES satellite in an active region AR 11692 on March 15$^{th}$ 2013. The flare was associated with a halo coronal mass ejection (CME). The flare created a global disturbance in the active region. {The CME associated with the flare produced a shock wave. To observe shock fronts, we created running difference images of AIA 193~\AA. The shock fronts are marked with red and green arrows in Figure~\ref{shock_fig}. It is evident from Figure~\ref{shock_fig} that the shock fronts are moving outward.}

\begin{figure}[ht!]
\centering
\includegraphics[scale=0.81,angle=90]{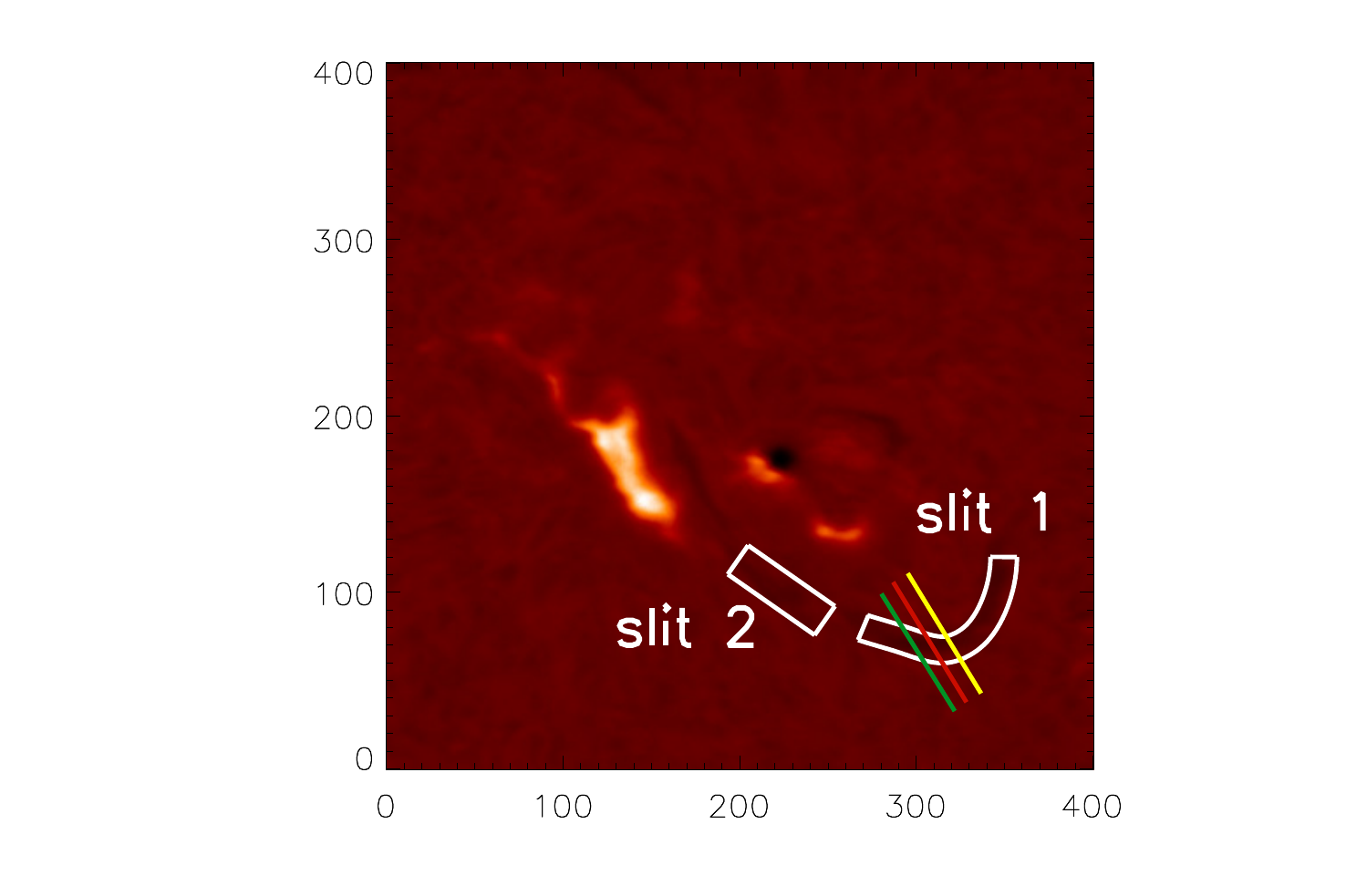}
\caption{NSO/GONG H$\alpha$ image of the filament lying at south-west of the active region AR 11692. Two broad artificial slices along the axis of the filament, used for generating x--t maps, are over plotted. Three artificial slits perpendicular to the axis of filament are overplotted in green, red and yellow, which were used in \cite{2015RAA....15.1713P} to investigate the transverse oscillations.}
\label{context}
\end{figure}

\begin{figure}[ht!]
\includegraphics[scale=0.6,angle=90]{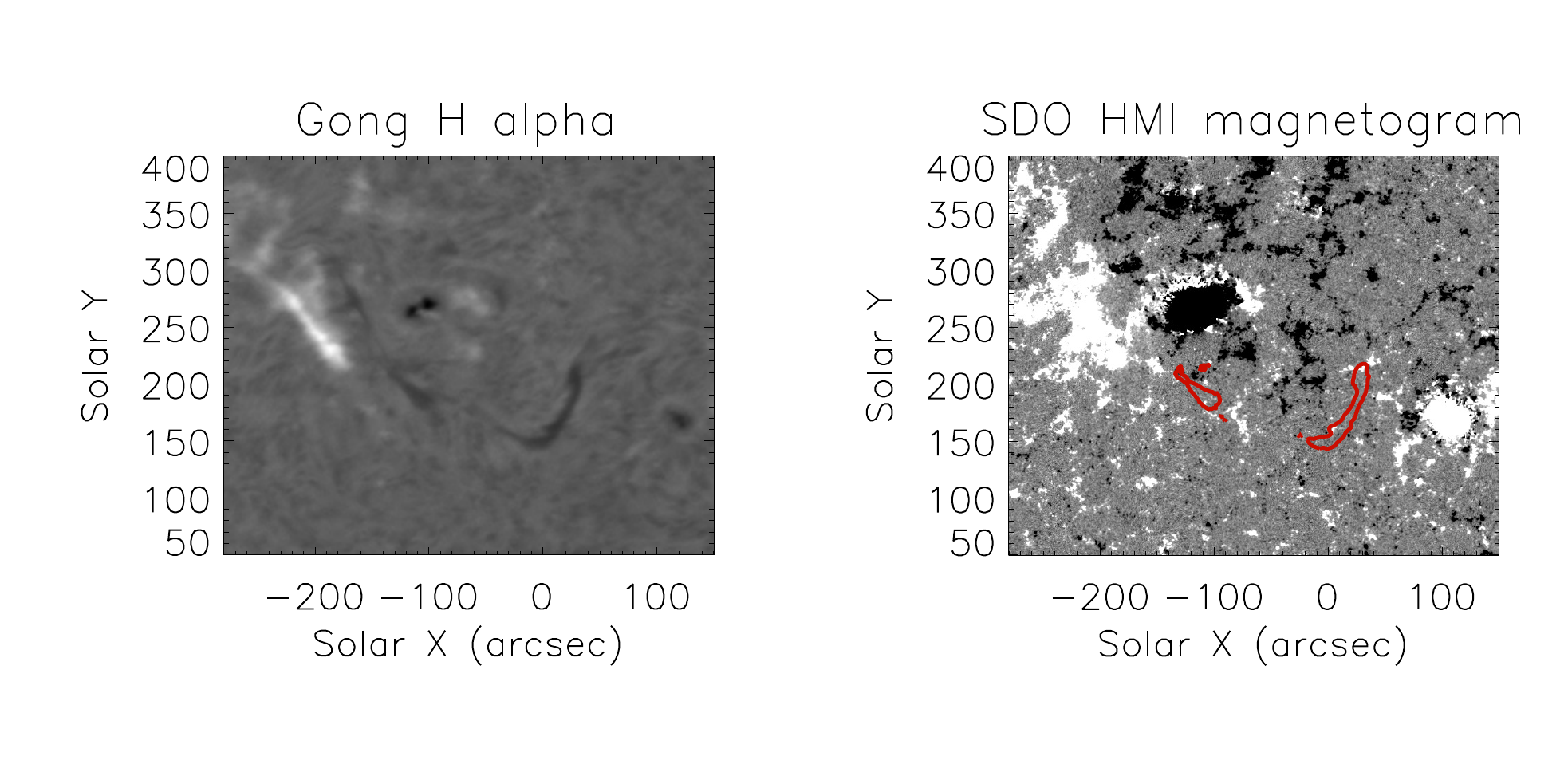}
\caption{{\it Left}: GONG H$\alpha$ image. {\it Right} HMI magnetogram. Overplotted red contours outline the position of the filament.}
\label{hmi}\vspace*{-1cm}
\end{figure}

A filament was lying to the south-west of the active region as seen in H$\alpha$ images from NSO/GONG  (Figure~\ref{context}).  NSO/GONG provides full-disk observation of the Sun in 6563~\AA ~(H$\alpha$). It has a pixel resolution of $\sim$ 1.07\arcsec and a cadence of 1 sec. The observational data used in this study are obtained from 06:00:54 UT to 10:22:54 UT. The flare  started at 05:46:00 UT, peaked at 06:58:00 UT and ended at 08:35:00 UT as recorded in GOES catalogue. We {note that just after the onset of the flare, oscillations were triggered in the filament}.\\
Filaments consist of fine threads that are not often seen in GONG H$\alpha$ images due to its poor spatial resolution. Therefore, we use the images from the extreme ultraviolet (EUV) passbands of \textit{Atmospheric Imaging Assembly} (AIA) onboard \textit{Solar Dynamic Observatory} (SDO). AIA instrument provides almost simultaneous full-disk images of the Sun in seven EUV band. It has a spatial resolution of $1.3''$, pixel size of $0.6''$ and a cadence of 12~s \citep{2012SoPh..275...17L}. {We find that the filament consists of several threads that are seen more clearly} in AIA 171~\AA~than any other EUV passbands of AIA. Therefore, we use AIA 171~\AA~ for further analysis. Moreover, we find that shortly after the oscillations started, the full disk images of the Sun were not available from 06:22:23 UT to 07:36:00 UT. Thus, the first cycle of oscillations is not seen in AIA 171~\AA. \\
We overplot the contours of the filament as seen in GONG H$\alpha$ over the magnetogram obtained from \textit{Helioseismic and Magnetic Imager} (HMI) onboard SDO (Figure~\ref{hmi}). {We find that the filament  lies over the} region which separates the positive and the negative polarity. We note from the movie (available online) that at 06:11:59 UT, {\it i.e} before eruption, the filament material as enclosed in two different contours are part of the same filament \cite[right panel of Figure~1] {2015RAA....15.1713P}. \\
By careful inspection of the H$\alpha$ images in the movie (available online), it appears that both longitudinal and transverse oscillations are present  simultaneously in the filament. {In the movie (available online),} we mark the filament in AIA images with a green arrow. The transverse oscillations in the filament under study are already reported in \inlinecite{2015RAA....15.1713P}. In this paper, we focus on longitudinal oscillations at two parts of the filament adjacent to its two {ends}.

\vspace*{-1.3cm}

%%%%%%%%%%%%%%%%%%%%%%%%%%%%%%%%%%%%%%%%%%%%%%%%%%%%%%%%%%%%%%%%%%%%%%%%%%%%%%%%%

\section{Data Analysis}
\subsection{GONG H$\alpha$}
   
To investigate the longitudinal oscillations in GONG H$\alpha$ images, we place two artificial broad slices, slit~1 and slit~2 along the axis of the filament as shown in Figure~\ref{context}. 
The length (width) of the $1^{st}$ and the $2^{nd}$ slice is 130 pixel (15 pixel) and 60 pixel (20 pixel) respectively. The filament is oscillating in both transverse and longitudinal directions simultaneously. We choose broad slices so that the filament material remain inside the slices in spite of the transverse oscillations. 

To further characterise the longitudinal oscillations in the filament, first we create a time-distance (x--t) map with time in x axis and distance along the slice in y axis for both slit~1 and slit~2. Intensity along each column of the x--t map is fitted with a Gaussian curve. The mean value and one sigma error of the intensity is estimated. Finally the x--t map is fitted with a damped sinusoidal function to investigate the oscillation characteristics. Since the filament is oscillating and the amplitude of the oscillations is decreasing with time, it can be fitted with a damped sinusoidal function (assuming the mass of the filament to be constant) and represented as
\citep{1999ApJ...520..880A},

\begin{equation}
y(t)=c+A_{sin}sin(\omega t+\phi_{sin})e^{-t/\tau_{sin}},
\end{equation}
where $c$ is a constant, $A_{sin}$ is the amplitude, $\tau_{sin}$ is the damping time, $\omega$ is the angular frequency, and $\phi_{sin}$ is the initial phase of sinusoidal function. 
The least square fitting is done using the function {\it MPFIT.pro} in Interactive Data Language (IDL) \citep{2009ASPC..411..251M}. 
{ It has been shown that if a filament accretes mass then the Bessel function describes the oscillations better than the damped sinusoid  \citep{2012ApJ...750L...1L}.}  
Therefore, we fit the x--t map using a damped zeroth order Bessel function of first kind, represented by
\begin{equation}
y(t)=c_{1}+A_{bes} J_{0} (\omega t+\phi_{bes})e^{-t/\tau_{bes}},
\end{equation}
where $c_{1}$ is a constant, $A_{bes}$ is amplitude, $\omega$ is the angular frequency, $\tau_{bes}$ is the damping time, and $\phi_{bes}$ is the phase of the Bessel function.

\subsection{AIA 171~\AA}
We repeat the analysis on AIA~171~\AA~ images but with only one artificial slice. The filament material at the position of slit~2 is not clearly seen in AIA~304, 171 and 193~\AA~ due to the presence of overlying post-flare loops (see movie).

\section{Results}
\subsection{GONG H$\alpha$}
The  period, damping time and amplitude estimated from the fitting of a damped sinusoidal function represented by Equation~(1) are summarized in Table~\ref{table1}. The  period, damping time and amplitude estimated from the fitting of a damped zeroth  order Bessel  function of first kind represented by Equation~(2) is summarized in Table~\ref{table2}. In the right panel of Figure~\ref{xt}, we overplot dashed green lines at three time instances. It is evident that both ends of the filament started oscillating in different phase. It is also worth noting that the oscillations started at different times in the two ends of the filament.
\subsection{AIA 171~\AA}
\begin{figure}[ht!]
\centering
\includegraphics[scale=0.81,angle=90]{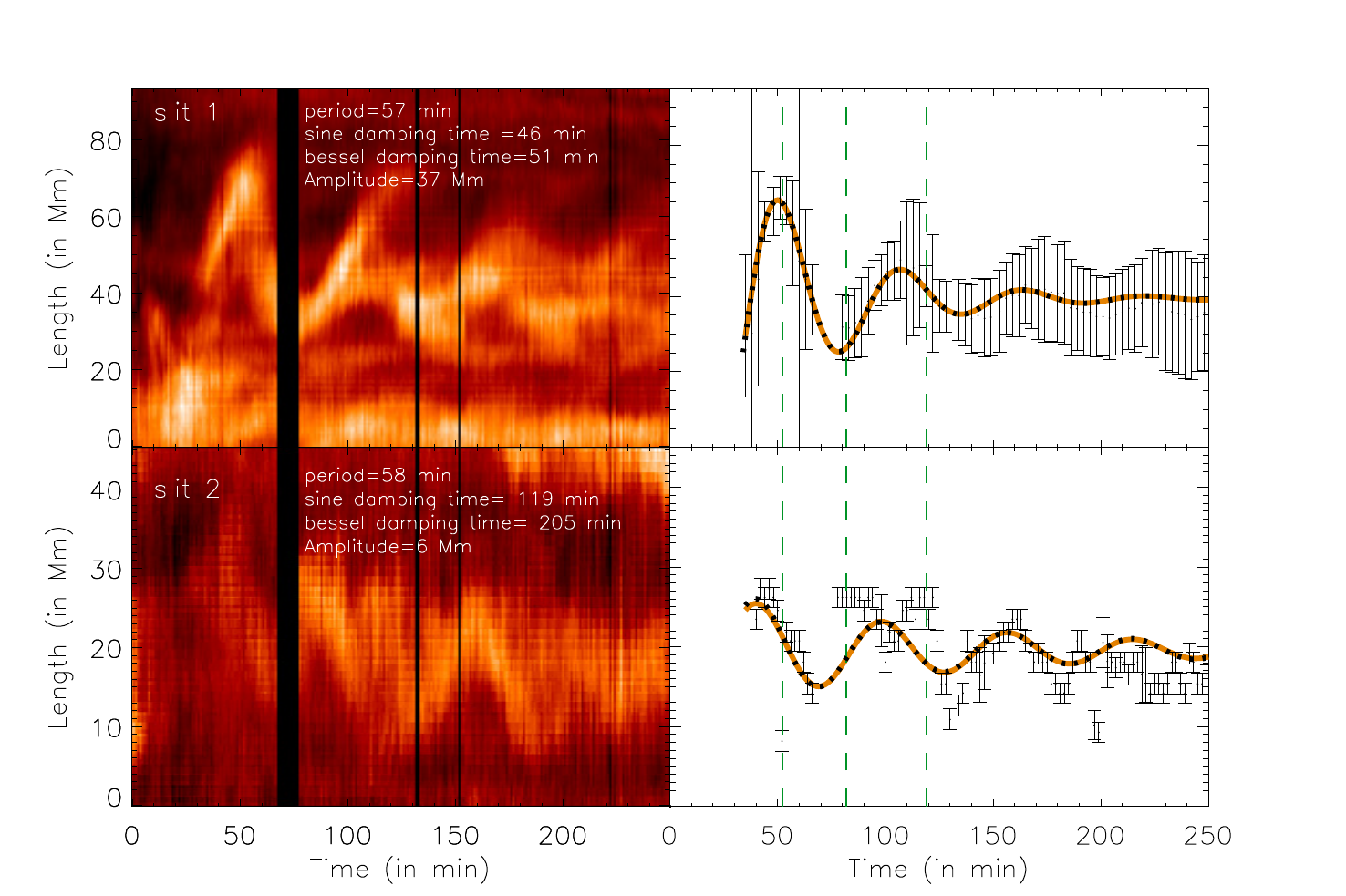}
\caption{Upper left panel shows the x--t map of slit~1. The thick black vertical strip represents the data gap between 07:07:54 Ut to 07:18:54 UT. The time period, amplitude and the damping time obtained from fitting is mentioned at the top of the plot. The lower left panel is same as upper left but for slit~2. In the upper right panel we plot the fitted damped sinusoidal curve with one sigma error bar. The lower right panel is same for the slice 2. Three vertical dashed lines over plotted at three time instances show that the oscillations started with different phases. The dotted black curve represents the best fit damped sinusoidal function while orange curve represents the damped Bessel function}
\label{xt}
\end{figure}

\begin{figure}[ht!]
\centering
\includegraphics[scale=0.81,angle=90]{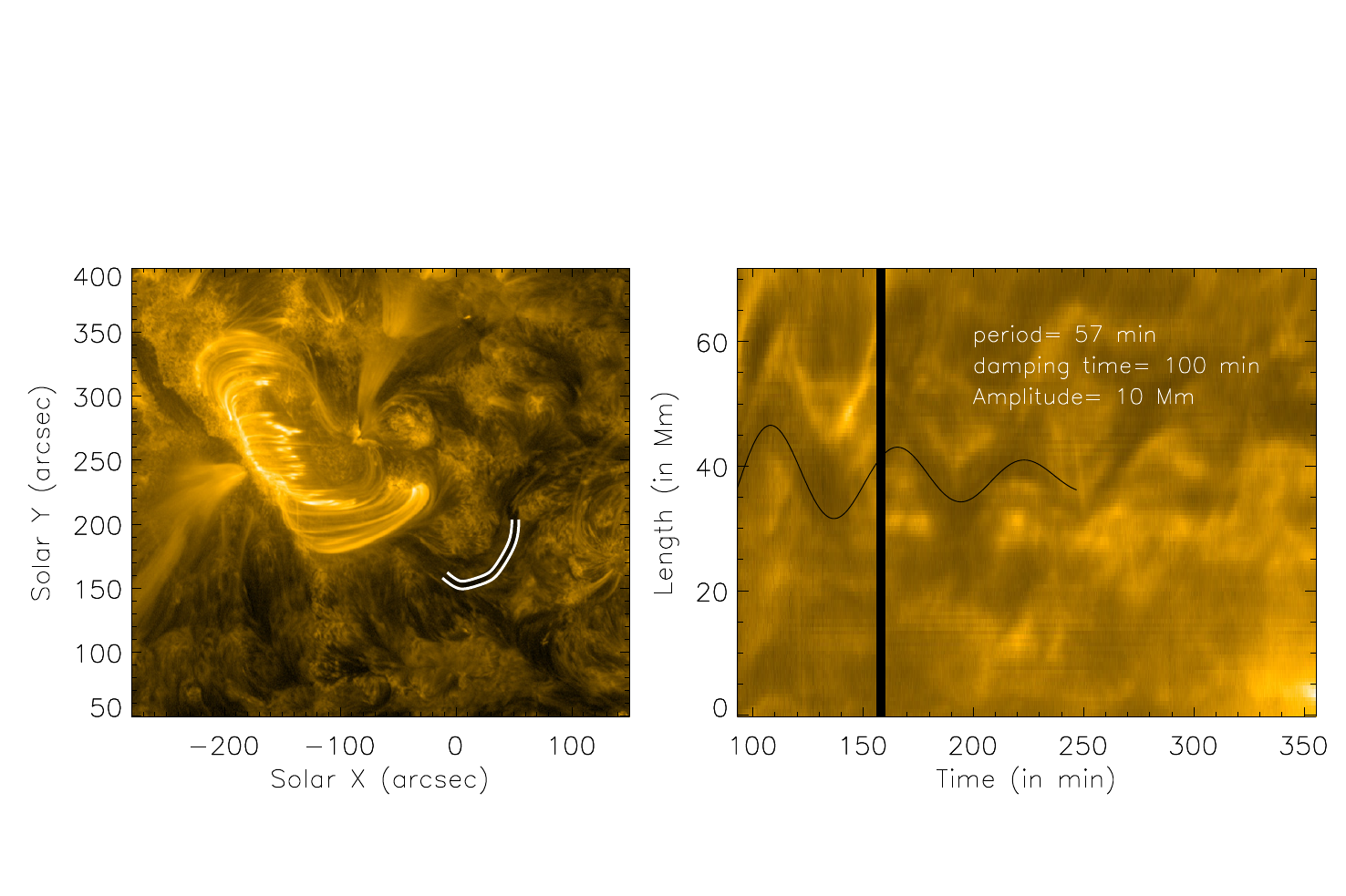}
\caption{The left panel shows the AIA~171~\AA~ image with artificial slice overplotted on the filament. The right panel represents the x--t map corresponding to the curve artificial slice shown in the left panel. The thick black vertical strip represents the data gap. Black curve represents the best fit sinusoidal function. The time period, amplitude and the damping time obtained from fitting is mentioned at the top.}
\label{xt_aia}
\end{figure}

 The period of oscillation is found to be 57$\pm$3 min which is similar to the period estimated at the position of slit~1 in GONG H$\alpha$. The damping time is estimated to be 100 min which is larger as compared to the damping time estimated using slit~1 in GONG H$\alpha$ images. Amplitude of oscillation is found to be 10~Mm which is less than the amplitude measured in GONG H$\alpha$ images. The main reason of large damping time and smaller amplitude is that, first cycle of the oscillations is missed in AIA~171~\AA~images, therefore the oscillations start after 100 min as shown in Figure~\ref{xt_aia} (right panel). We note from Figure~\ref{xt} (top right panel) that the {estimated damping time} of oscillations and amplitude are strongly dependent on first cycle of the oscillations. If first cycle of oscillations is missed then the damping time will be longer and amplitude will be smaller. {Another reason can be that the filament threads apparently merge and separate during the oscillations. This may be attributed to the fact that multiple threads are stacked along the line of sight, so if they are not moving in phase it may appear that they are merging and separating. This is probably also related to the presence of transverse oscillations because such effect is not reported in earlier studies \citep{2012ApJ...750L...1L} where only longitudinal oscillations were present. Therefore, there is large uncertainty in choosing the thread that is associated with the oscillations (see right panel of Figure~\ref{xt_aia}). It is worth noting at this point that we did not fit damped Bessel function to AIA dataset because, as said above, the first cycle of oscillation is missed thus the Bessel function will not give the correct estimation of the damping time.}

%%%%%%%%%%%%%%%%%%%%%%%%%%%%%%%%%%%%%%%%%%%
%%%%%%%%%%%%%%%%%%%%%%%%%%%%%%%%%%%%%%%%%%%

\subsection{Radius of curvature of magnetic dip}
Filament material is supported by the magnetic fields against gravity. We use a simple pendulum model to estimate the oscillation parameters \citep{2012ApJ...750L...1L}. Assuming restoring force to be gravity, we get the following equation
\begin{equation}
\omega=\frac{2\pi}{P}=\sqrt{\frac{g_{0}}{R}},
\end{equation}
where $g_{0}$ is the Sun's surface gravity, $\omega$ is the oscillation frequency, $P$ is the period of oscillation and $R$ is the radius of curvature.
%\citep{2012ApJ...757...98L}.
{Using Equation (3), the radius of curvature in GONG H$\alpha$ images is estimated to be $\sim$ 80 Mm and 85 Mm at the position of dip 1 (where $1^{st}$ slice is placed) and dip 2 (where $2^{nd}$ slice is placed; see, Figure~\ref{context}) respectively. 
Similarly, the radius of curvature at the location of dip 1 (see left hand panel of Figure~\ref{xt_aia} where artificial slice is placed) in AIA~171~\AA~is estimated to be 80 Mm. }
%The radius of curvature is a bit larger than the radius of curvature estimated (43-66 Mm) by \inlinecite{2014ApJ...785...79L}.
\subsection{Estimation of magnetic field strength}
We calculate a lower limit of the magnetic field strength, assuming that magnetic stress is balanced  by the weight of the filament material present in a dip \citep{2012ApJ...750L...1L}, using the following expression, 
\begin{equation}
B(G) \geq 26 \left( \frac{n_{e}}{10^{11}}\right) ^\frac{1}{2}  P ,
\end{equation}
where $B$ is the magnetic field, $n_{e}$ is the electron number density in cm$^{-3}$ and $P$ is the period of oscillation in hours. Here we use the typical value of electron number density $10^{11}$ cm$^{-3}$ as reported in \inlinecite{2010SSRv..151..243L}. Therefore, minimum magnetic field strength is estimated to be $\sim$ (25 $\pm$ 1) G both in dip 1 and dip 2 which is consistent with the typical values of the measured magnetic fields in the filaments \citep{2010SSRv..151..243L, 2010SSRv..151..333M}.  Since the period of oscillation is estimated to be the same using AIA~171~\AA, the {estimated magnetic field strength remains unchanged.}

\subsection{Estimation of the Mass Accretion Parameter}
In the model given by \inlinecite{2012ApJ...750L...1L}, the damping of longitudinal oscillations in threads of the filaments is explained by the accretion of filament material. We calculate the mass accretion rate from the best fit parameters of the damped Bessel function { (represented by Equation~(2))}, using the following relation \citep{2014ApJ...785...79L},
$$\alpha=\frac{\omega m_{0}}{\phi_{bes}},$$
where, $m_{0}$ is the initial mass of filament, $\omega$ and $\phi_{bes}$ are the obtained from fitting of the damped Bessel function. The mass of a filament, $m_{0}$ can be calculated as $m_{0}=1.27 m_{p} n_{e} \pi r^2 L$, where $r$ and $L$ are the radius and the length of the thread of the filament. We take the typical radius of a filament thread to be 100 km as assumed in \inlinecite{2014ApJ...785...79L}. {The length of the thread is estimated using x--t map as shown in Figure~\ref{xt} (top panel). As explained above, the artificial slit is placed along the length of the filament. Thus the extent of the brightness along each column of the x--t map gives an estimate of the length of the filament thread at a particular time instance.  The length of the filament thread is measured at various time instances ({\it i.e,} along different columns) from the start to the end of the oscillations. Finally, we estimate the mean value of the length of the filament thread $\sim$ 23$\pm$ 7 Mm at the position of slit 1. Following the above procedure, the length of the filament thread at the position of slit 2 is estimated $\sim$  10$\pm$4 Mm. {It should be borne in mind that the estimation of the length of the filament thread at the position of slit 2 is uncertain because the oscillations are not clearly seen in the x--t maps as shown in the bottom panel of Figure~\ref{xt}.} Under these assumptions we estimated the mass accretion rate at the position of slit 1 and 2 to be $\sim$ (51 $\pm$ 17 $\times $10$^{6}$ kg hr$^{-1}$) and (192 $\pm$ 72 $ \times $10$^{6}$ kg hr$^{-1}$) respectively.} {The derived mass accretion rate at the position of slit~1 is comparable to those reported in earlier studies  \citep{2014ApJ...785...79L}.} It should be noted that at the position of slit 2 the mass accretion rate is higher while damping time is longer. One of the reasons for this discrepancy is the uncertainty in the fitting parameters. Since the mass accretion rate depends on the phase of the Bessel function which is estimated from the best fit parameters. Therefore, if the fitting of a function is not good then the value of phase will have large errors and thus can not be trusted.
% We did not use AIA data to estimate the mass accretion rate because in AIA data the first cycle of oscillation is completely missed thus fitting Bessel function may not give correct values for mass accretion rate. 

\begin{table}
\caption{Table of parameters of damped sine fitting}
\label{table1}
\begin{tabular}{ccccc}     % define the column alignment
                           % l: left, c: center, r: right
  \hline                   % horizontal line
  Slit No. & Displacement amplitude & Period & Damping time & Velocity amplitude \\
        & (in Mm) & (in min) & (in min) & (in km s$^{-1}$)\\
  \hline
\textbf{ slice 1} &   \textbf{37}$\pm$ \textbf{15}  & \textbf{57}$\pm$\textbf{2} & \textbf{46}$\pm$\textbf{8} & \textbf{69}$\pm$\textbf{30}  \\
\textbf{slice 2} & 
\textbf{6} $\pm$\textbf{1}  & \textbf{58}$\pm$\textbf{1} & \textbf{119.3}$\pm$\textbf{0.1}&  \textbf{15}$\pm$\textbf{2}\\

  \hline
\end{tabular}
\end{table}

\begin{table}
\caption{Table of parameters of modified Bessel function fitting}
\label{table2}
\begin{tabular}{ccccc}     % define the column alignment
                           % l: left, c: center, r: right
  \hline                   % horizontal line
  Slit No. & Displacement amplitude & Period & Damping time & Velocity amplitude \\
        & (in Mm) & (in min) & (in min) & (in km s$^{-1}$)\\
  \hline
\textbf{ slice 1} &   \textbf{27}$\pm$\textbf{12}  & \textbf{57}$\pm$\textbf{2} & \textbf{51}$\pm$\textbf{10} & \textbf{49}$\pm$\textbf{25}  \\
\textbf{slice 2} & \textbf{6}$\pm$\textbf{2}  & \textbf{58}$\pm$\textbf{1} & \textbf{204}$\pm$\textbf{38}&  \textbf{11}$\pm$\textbf{4}\\

  \hline
\end{tabular}
\end{table}
%--------
\subsection{Possibility of existence of both longitudinal and transverse wave from kink oscillation}
In this subsection we  investigate the possibility of the existence of both longitudinal and transverse oscillations driven by kink oscillations. We assume the filament as a cylinder embedded in a uniform plasma with low plasma $\beta$. Magnetic field is assumed to be constant inside the cylinder and the effects of gravity is ignored.
The model set up is similar to the one described in \cite{2016ApJS..223...23Y}. To get the variation of thermodynamic quantities, linearized ideal MHD equations are used \citep[see][]{2009SSRv..149..199R},
\begin{equation}
\rho=- \nabla  .  (\rho _{0}   \xi),
\end{equation}

\begin{equation}
\rho_{0} \frac{\partial^{2} \mathbf{ \xi } }{\partial t^{2}}=-\nabla (p + \mathbf{b.B_{0}/ \mu_{0}}) + \frac{1}{\mu} [(\mathbf{B_{0}}.\nabla)\mathbf{b} + (\mathbf{b}.\nabla)\mathbf{B_{0}}],
\end{equation}

\begin{equation}
\mathbf{b}=\nabla \times(\bf{ \xi} \times \mathbf{B_{0}}),
\end{equation}

\begin{equation}
 p - C_{s} ^{2} \rho = \mathbf{\xi}  .  (C_{s} ^{2} \nabla \rho_{0} - \nabla p_{0}),
\end{equation}
\noindent
where $\xi$ is the Lagrange displacement vector, $\rho$, $p$ and {\bf b} are perturbed quantities while $\rho_{0}$, $p_{0}$ and {\bf B$_{0}$} are unperturbed quantities. {Neumann boundary conditions are assumed at $r=a$
 $$[P_{T}]_{r=a}=0,$$ $$[\xi_{r}]_{r=a}=0$$ and Dirichlet boundary conditions are assumed at $r=0,\infty$
$$P_{T}|_{r=0}<\infty,$$ $$ \xi^{2}|_{r=0}=\infty,$$
$$P_{T}|_{r\rightarrow \infty}=0,$$ $$\xi|_{r\rightarrow \infty}=0,$$
where $P_{T}$ is the total pressure.}

%\citep{2015ApJ...807...98Y}
%\citep{2013A&A...555A..74A}
For a filament, we define  $\omega$, $\omega_{A}$, $\omega_{T}$ as  kink, Alfv\'en and tube oscillation frequency respectively and the tube speed, $C_{T}$ is given by
 $C_{T}=\frac{C_{A}C_{s}}{\sqrt{C_{A}^{2}+C_{s}^{2}}}$ where $C_{A}=\frac{B_{0}}{\sqrt{\mu_{0}\rho_{0}}}$ is the Alfv\'en speed  and 
 $C_{s}=\sqrt{ \frac{\gamma p_{0}}{\rho_{0}}}$ is the sound speed.
 $B_{0}$,  $p_{0}$, $\rho_{0}$ are equilibrium magnetic filed, plasma pressure and plasma density respectively and $\mu_{0}$ is the magnetic permeability in free space. 
 For kink mode (m=1), the total pressure perturbation, $P_{T1}=p + \mathbf{b.B_{0}/ \mu_{0}}$, in the filament is assumed to be 
\begin{equation}
P_{T1} = A R(r) \cos (\omega t) \sin (kz) \cos (\phi) 
\end{equation}
 where $A$ is the amplitude of the total perturbed pressure. Since kink mode is not axisymmetric thus cos($\phi$) dependence is assumed. Substituting of Equation (9) into Equations~(5-8) yields,
 \begin{equation}
\frac{d^{2}P_{T1}}{dr^{2}}+\frac{dP_{T1}}{rdr}-(\kappa^{2}_{r}+\frac{1}{r^{2}})P_{T1}=0,
\end{equation}
where 
\begin{equation}
\kappa^{2}_{r} = \frac{(\omega_{s}^{2}-\omega^{2})(\omega_{A}^{2}-\omega^{2})}{(\omega_{s}^{2}+\omega_{A}^{2})(\omega_{T}^{2}-\omega^{2})}k^{2},
\end{equation}
 where $\omega_{s}^2=k^2 C_{s}^2$.  
{ Equation~(10) should also hold true for R(r). In this work we confine ourselves to the inside of the flux tube thus, on solving Equation~(10) for R(r), we get,
$$R=J_1(|\kappa_{r}|r) .$$ }
%and therefore
%$$P_{T1} = A J_1(|\kappa_{r}|r) \cos (\omega t) \sin (kz) \cos (\phi).  $$}
In cylindrical coordinates, $v_{r}$ is the perturbed velocity amplitude of the transverse oscillation and $v_{z}$ is the perturbed velocity amplitude of the longitudinal oscillation. { Inside the cylinder  the perturbed radial and axial component of velocity amplitude} are defined as \citep[see][for details]{2016ApJS..223...23Y},
$$v_{r}=-\frac{rdR}{Rdr} \sin(\omega t) \sin(kz) \cos(\phi)$$
and $$v_{z} = - \frac{C_{T}^{2}}{C_{A}^{2}}kr\frac{\omega^{2}-\omega_{A}^{2}}{\omega^{2}-\omega_{T}^{2}} \sin(\omega t) \cos(kz) \cos(\phi).$$
Thus, axial to radial velocity amplitude ratio is given by the following expression,

\begin{equation}
\frac{v_{z}}{v_{r}}=k\frac{C_{T}^{2}}{C_{A}^{2}}\frac{R}{\frac{dR}{dr}}\frac{\omega^{2}-\omega_{A}^{2}}{\omega^{2}-\omega_{T}^{2}}cot(kz) .
\end{equation}
 We take the length of slit~2 as length of the filament, which is estimated to be 90 Mm. The radius of the filament is estimated to be $\sim$ 1 Mm. {It is important to note that here we assume the length of the flux tube (modelled as a filament) as the length of the slit covering the filament material because here we compare longitudinal oscillations with the transverse oscillations as reported by \inlinecite{2015RAA....15.1713P}, where authors assumed that filament was oscillating as a whole in the transverse direction.} To estimate the ratio of the axial to radial velocity amplitude, we choose three slit positions as shown in green, red and yellow lines in Figure~\ref{context}, that were used by  \inlinecite{2015RAA....15.1713P}. From observations, we find that the longitudinal velocity amplitude to the transverse velocity amplitude ratio at these locations are 3.92, 3.26 and 2.22 respectively. Using {Equation (12)}, we estimate the ratio to be 4 $\cdot$ $10^{-4}$, 2 $\cdot$ $10^{-4}$ and 1 $\cdot$ $10^{-5}$ respectively assuming typical value of sound speed in chromosphere as 15 km s$^{-1}$,  Alfv\'en speed in three slit positions as 82, 88 and 91 km s$^{-1}$ respectively and perturbation frequency (assumed to be equal to the kink oscillation frequency) to be 117, 124 and 128 km s$^{-1}$ respectively \citep[see,][]{2015RAA....15.1713P}. It should be borne in mind that a linear analysis is presented here. Considering the {observed displacement} to be large, non-linear effects would become important. Therefore to model such events further studies are needed. Nevertheless, from above analysis, we note that it is almost impossible to detect longitudinal oscillation in low-$\beta$ plasma as a result of fast kink mode in our observation. {Therefore, large amplitude longitudinal oscillations cannot be the kink mode.}
 
 \section{Interaction of shock with filament}
In this section, we explore the {most likely mechanism for driving the} longitudinal and transverse oscillations in a filament channel.  
 \inlinecite{2014ApJ...795..130S} reported that shock waves can interact with filaments in two possible geometries. If shock waves come such that the normal vector to the shock wave is perpendicular to the filament axis then a filament oscillates in the transverse direction. While if the shock wave interacts with a filament such that the normal vector of the shock wave is parallel to the filament axis then a filament oscillates in the longitudinal direction. Here we are proposing a more general scenario where a shock wave interacts obliquely with the filament such that the filament material oscillates in both transverse and longitudinal direction as {revealed} from our observations. {The position of the shock fronts in our observations are clearly seen in Figure~\ref{shock_fig}. }However, it should be borne in mind that from observations it is not straight forward to {determine} the direction of shock with respect to the filament because the filament is in close proximity of the active region and is associated with overlying loops that expand in response to the shock wave. This makes it extremely difficult to {determine} the orientation of the shock wave. 

\begin{figure}[htbp!]
\center{\includegraphics[scale=0.4,clip=]{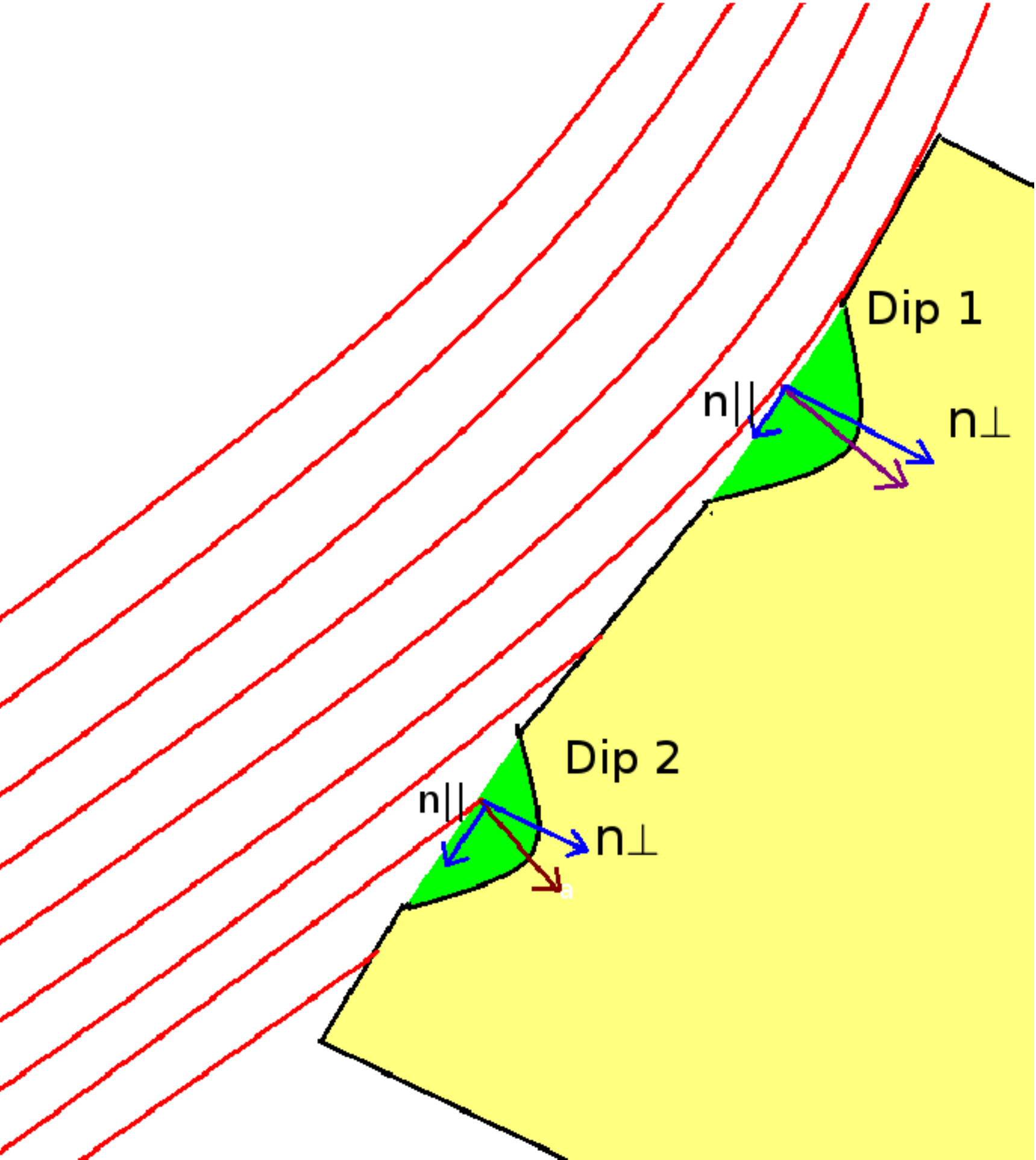}}
\caption{A cartoon depicting the interaction of a shock wave with a filament. The filament has two magnetic dips, Dip 1 and Dip 2. The filament material is shown in green. The shock wave front is represented by red curves. The normal vector to shock wave front is shown in dark red. {Arrows in blue represent the parallel and perpendicular component of the normal vector.}}
\label{cartoon}
\end{figure}

We depict this by a cartoon in Figure~\ref{cartoon}. We show that when a shock wave hit a filament obliquely, the normal vector to the wave front, $\hat{n}$ (shown in dark red), can be decomposed into perpendicular ($n_{\bot}$) and parallel ($n_{||}$) components as shown in Figure~\ref{cartoon} in blue. The component parallel to the filament axis triggers the longitudinal oscillations in the filament while the component perpendicular to the filament axis triggers the transverse oscillations in the filament.
Thus the shock front interacts with the filament at dip 1 and displaces the filament material in both transverse and longitudinal direction. Similarly in dip 2 the shock front displaces the plasma in both transverse and longitudinal directions, but at a different time. With this schematic picture we can explain the different start time of longitudinal oscillation at two ends of the filament. However,  with this scenario, one should expect the transverse oscillations to be present at the locations of both dips but transverse oscillations are only observed at the location of dip 1. It is worth noting that this is the simple geometry. In principle many parts of the shock wave can interact with the filament {at different orientations}  and can excite different modes. Further studies are needed to confirm this.

\section{Summary and Conclusions}
In this work, we find the co-existence of longitudinal and transverse oscillations in an active region filament as observed by NSO/GONG H$\alpha$. The oscillations are triggered by the interaction of a shock wave with the filament. We find that  the oscillations in two ends of the filament started with different phase but with almost similar period of oscillations. We note that the longitudinal oscillations at two slit positions started at different times (see, Figure~\ref{xt}) may be because the shock wave front interacts with the two parts of filament at different time instances. We depict it with a schematic diagram (see, Figure~\ref{cartoon}).

The damping time at the position of slit 1 is estimated to be 46 minute. This is a bit too strong as compared to the earlier reports \citep{2009SSRv..149..283T} and comparable to what is reported in \inlinecite{2014ApJ...785...79L}. \inlinecite{2012ApJ...750L...1L} suggested that the mass accretion of the filament material is responsible for damping. We find mass accretion rate to be 51 $\pm$ 17 $\times $10$^{6}$ kg hr$^{-1}$ at the position of slit 1 and 192 $\pm$ 72 $ \times $10$^{6}$ kg hr$^{-1}$ at the position of slit 2 {which is comparable with the earlier reports \citep{2014ApJ...785...79L}.}  We performed the filament seismology and estimated the magnetic field strength at two dip locations. We estimated the lower limit of the magnetic field strength $\sim$ (25 $\pm$ 1) G at the location of both the slices which is consistent with typical values of measured magnetic field from observation \citep{2010SSRv..151..333M}. We also calculate the radius of curvature of two magnetic dips of the filament as 80 Mm and 85 Mm for dip 1 and dip 2, respectively.\\

From our observations we exclude the possibility of existence of transverse and longitudinal oscillation simultaneously driven by a fast kink mode. Longitudinal oscillations due to kink wave may be present but their amplitude is far less as compared to the amplitude of the longitudinal oscillations observed in this case.  Thus we believe that transverse and longitudinal oscillations exist independently in the filament under study. 
 \acknowledgement
We would like to thank the referee for his/her valuable comments which has improved the presentation of this article. 
%%% BIBLIOGRAPHY %%%%%%%%%%%%%%%%%%%%%%%%%%%%%%%%%%%%%%%%%%%%%%%%%%%%%%%%%%%

  % format of references provided by the journal (.bst)
%\bibliographystyle{spr-mp-sola}
%\bibliographystyle{spr-mp-sola}
     % name your Bibtex file containing your references (.bib)
%\bibliography{reference_filament.bib}  

\begin{thebibliography}{34}
% BibTex style file: spr-mp-sola.bst (nameyear), 2015-03-09
\ifx\bisbn     \undefined \def\bisbn  #1{ISBN #1}\fi
\ifx\binits    \undefined \def\binits#1{#1}\fi
\ifx\bauthor   \undefined \def\bauthor#1{#1}\fi
\ifx\batitle   \undefined \def\batitle#1{#1}\fi
\ifx\bjtitle   \undefined \def\bjtitle#1{\textit{#1}}\fi
\ifx\bvolume   \undefined \def\bvolume#1{\textbf{#1}}\fi
\ifx\byear     \undefined \def\byear#1{#1}\fi
\ifx\bissue    \undefined \def\bissue#1{#1}\fi
\ifx\bfpage    \undefined \def\bfpage#1{#1}\fi
\ifx\blpage    \undefined \def\blpage #1{#1}\fi
\ifx\burl      \undefined \def\burl#1{\textsf{#1}}\fi
\ifx\href      \undefined \def\href#1#2{\textsf{#2}}\fi
\ifx\betal     \undefined \def\betal{\textit{et al.}}\fi
\ifx\bctitle   \undefined \def\bctitle#1{#1}\fi
\ifx\beditor   \undefined \def\beditor#1{#1}\fi
\ifx\bbtitle   \undefined \def\bbtitle#1{\textit{#1}}\fi
\ifx\bedition  \undefined \def\bedition#1{#1}\fi
\ifx\bseriesno \undefined \def\bseriesno#1{\textbf{#1}}\fi
\ifx\blocation \undefined \def\blocation#1{#1}\fi
\ifx\bsertitle \undefined \def\bsertitle#1{\textit{#1}}\fi
\ifx\bsnm      \undefined \def\bsnm#1{#1}\fi
\ifx\bsuffix   \undefined \def\bsuffix#1{#1}\fi
\ifx\bparticle \undefined \def\bparticle#1{#1}\fi
\ifx\barticle  \undefined \def\barticle#1{}\fi
\ifx\binstitute  \undefined \def\binstitute#1{#1}\fi
\ifx\bpublisher  \undefined \def\bpublisher#1{#1}\fi
\ifx\doiurl    \undefined
  \def\doiurl#1{\href{http://dx.doi.org/#1}{\textsf{DOI}}}\fi
\ifx\arxivurl  \undefined
  \def\arxivurl#1{\href{http://arxiv.org/abs/#1}{\textsf{arXiv}}}\fi
\ifx\adsurl    \undefined
  \def\adsurl#1{\href{http://adsabs.harvard.edu/abs/#1}{\textsf{ADS}}}\fi
\ifx\botherref \undefined \def\botherref#1{}\fi
\ifx\url       \undefined \def\url#1{\textsf{#1}}\fi
\ifx\bchapter  \undefined \def\bchapter#1{}\fi
\ifx\bbook     \undefined \def\bbook#1{}\fi
\ifx\bcomment  \undefined \def\bcomment#1{#1}\fi
\ifx\oauthor   \undefined \def\oauthor#1{#1}\fi
\ifx\citeauthoryear \undefined\def \citeauthoryear#1{#1}\fi
\ifx\endbibitem\undefined \def\endbibitem{}\fi
\ifx\bconflocation  \undefined \def\bconflocation#1{#1} \fi

\bibitem[\protect\citeauthoryear{{Andries}, {Arregui}, and
  {Goossens}}{2005}]{2005ApJ...624L..57A}
\begin{barticle}
\bauthor{\bsnm{{Andries}}, \binits{J.}},
\bauthor{\bsnm{{Arregui}}, \binits{I.}},
\bauthor{\bsnm{{Goossens}}, \binits{M.}}:
\byear{2005},
\batitle{{Determination of the Coronal Density Stratification from the
  Observation of Harmonic Coronal Loop Oscillations}}.
\bjtitle{\apjl}
\bvolume{624},
\bfpage{L57}.
\doiurl{10.1086/430347}.
\adsurl{2005ApJ...624L..57A}.
\end{barticle}
\endbibitem

\bibitem[\protect\citeauthoryear{{Arregui}, {Oliver}, and
  {Ballester}}{2012}]{2012LRSP....9....2A}
\begin{barticle}
\bauthor{\bsnm{{Arregui}}, \binits{I.}},
\bauthor{\bsnm{{Oliver}}, \binits{R.}},
\bauthor{\bsnm{{Ballester}}, \binits{J.L.}}:
\byear{2012},
\batitle{{Prominence Oscillations}}.
\bjtitle{Living Reviews in Solar Physics}
\bvolume{9},
\bfpage{2}.
\doiurl{10.12942/lrsp-2012-2}.
\adsurl{2012LRSP....9....2A}.
\end{barticle}
\endbibitem

\bibitem[\protect\citeauthoryear{{Arregui}
  \textit{et~al.}}{2007}]{2007A&A...463..333A}
\begin{barticle}
\bauthor{\bsnm{{Arregui}}, \binits{I.}},
\bauthor{\bsnm{{Andries}}, \binits{J.}},
\bauthor{\bsnm{{Van Doorsselaere}}, \binits{T.}},
\bauthor{\bsnm{{Goossens}}, \binits{M.}},
\bauthor{\bsnm{{Poedts}}, \binits{S.}}:
\byear{2007},
\batitle{{MHD seismology of coronal loops using the period and damping of
  quasi-mode kink oscillations}}.
\bjtitle{\aap}
\bvolume{463},
\bfpage{333}.
\doiurl{10.1051/0004-6361:20065863}.
\adsurl{2007A\%26A...463..333A}.
\end{barticle}
\endbibitem

\bibitem[\protect\citeauthoryear{{Arregui}
  \textit{et~al.}}{2008}]{2008ApJ...682L.141A}
\begin{barticle}
\bauthor{\bsnm{{Arregui}}, \binits{I.}},
\bauthor{\bsnm{{Terradas}}, \binits{J.}},
\bauthor{\bsnm{{Oliver}}, \binits{R.}},
\bauthor{\bsnm{{Ballester}}, \binits{J.L.}}:
\byear{2008},
\batitle{{Damping of Fast Magnetohydrodynamic Oscillations in Quiescent
  Filament Threads}}.
\bjtitle{\apjl}
\bvolume{682},
\bfpage{L141}.
\doiurl{10.1086/591081}.
\adsurl{2008ApJ...682L.141A}.
\end{barticle}
\endbibitem

\bibitem[\protect\citeauthoryear{{Aschwanden}
  \textit{et~al.}}{1999}]{1999ApJ...520..880A}
\begin{barticle}
\bauthor{\bsnm{{Aschwanden}}, \binits{M.J.}},
\bauthor{\bsnm{{Fletcher}}, \binits{L.}},
\bauthor{\bsnm{{Schrijver}}, \binits{C.J.}},
\bauthor{\bsnm{{Alexander}}, \binits{D.}}:
\byear{1999},
\batitle{{Coronal Loop Oscillations Observed with the Transition Region and
  Coronal Explorer}}.
\bjtitle{\apj}
\bvolume{520},
\bfpage{880}.
\doiurl{10.1086/307502}.
\adsurl{1999ApJ...520..880A}.
\end{barticle}
\endbibitem

\bibitem[\protect\citeauthoryear{{Gilbert}
  \textit{et~al.}}{2008}]{2008ApJ...685..629G}
\begin{barticle}
\bauthor{\bsnm{{Gilbert}}, \binits{H.R.}},
\bauthor{\bsnm{{Daou}}, \binits{A.G.}},
\bauthor{\bsnm{{Young}}, \binits{D.}},
\bauthor{\bsnm{{Tripathi}}, \binits{D.}},
\bauthor{\bsnm{{Alexander}}, \binits{D.}}:
\byear{2008},
\batitle{{The Filament-Moreton Wave Interaction of 2006 December 6}}.
\bjtitle{\apj}
\bvolume{685},
\bfpage{629}.
\doiurl{10.1086/590545}.
\adsurl{2008ApJ...685..629G}.
\end{barticle}
\endbibitem

\bibitem[\protect\citeauthoryear{{Goossens}, {Andries}, and
  {Arregui}}{2006}]{2006RSPTA.364..433G}
\begin{barticle}
\bauthor{\bsnm{{Goossens}}, \binits{M.}},
\bauthor{\bsnm{{Andries}}, \binits{J.}},
\bauthor{\bsnm{{Arregui}}, \binits{I.}}:
\byear{2006},
\batitle{{Damping of magnetohydrodynamic waves by resonant absorption in the
  solar atmosphere}}.
\bjtitle{Philosophical Transactions of the Royal Society of London Series A}
\bvolume{364},
\bfpage{433}.
\doiurl{10.1098/rsta.2005.1708}.
\adsurl{2006RSPTA.364..433G}.
\end{barticle}
\endbibitem

\bibitem[\protect\citeauthoryear{{Hershaw}
  \textit{et~al.}}{2011}]{2011A&A...531A..53H}
\begin{barticle}
\bauthor{\bsnm{{Hershaw}}, \binits{J.}},
\bauthor{\bsnm{{Foullon}}, \binits{C.}},
\bauthor{\bsnm{{Nakariakov}}, \binits{V.M.}},
\bauthor{\bsnm{{Verwichte}}, \binits{E.}}:
\byear{2011},
\batitle{{Damped large amplitude transverse oscillations in an EUV solar
  prominence, triggered by large-scale transient coronal waves}}.
\bjtitle{\aap}
\bvolume{531},
\bfpage{A53}.
\doiurl{10.1051/0004-6361/201116750}.
\adsurl{2011A\%26A...531A..53H}.
\end{barticle}
\endbibitem

\bibitem[\protect\citeauthoryear{{Isobe} and
  {Tripathi}}{2006}]{2006A&A...449L..17I}
\begin{barticle}
\bauthor{\bsnm{{Isobe}}, \binits{H.}},
\bauthor{\bsnm{{Tripathi}}, \binits{D.}}:
\byear{2006},
\batitle{{Large amplitude oscillation of a polar crown filament in the
  pre-eruption phase}}.
\bjtitle{\aap}
\bvolume{449},
\bfpage{L17}.
\doiurl{10.1051/0004-6361:20064942}.
\adsurl{2006A\%26A...449L..17I}.
\end{barticle}
\endbibitem

\bibitem[\protect\citeauthoryear{{Jing}
  \textit{et~al.}}{2003}]{2003ApJ...584L.103J}
\begin{barticle}
\bauthor{\bsnm{{Jing}}, \binits{J.}},
\bauthor{\bsnm{{Lee}}, \binits{J.}},
\bauthor{\bsnm{{Spirock}}, \binits{T.J.}},
\bauthor{\bsnm{{Xu}}, \binits{Y.}},
\bauthor{\bsnm{{Wang}}, \binits{H.}},
\bauthor{\bsnm{{Choe}}, \binits{G.S.}}:
\byear{2003},
\batitle{{Periodic Motion along a Solar Filament Initiated by a Subflare}}.
\bjtitle{\apjl}
\bvolume{584},
\bfpage{L103}.
\doiurl{10.1086/373886}.
\adsurl{2003ApJ...584L.103J}.
\end{barticle}
\endbibitem

\bibitem[\protect\citeauthoryear{{Jing}
  \textit{et~al.}}{2006}]{2006SoPh..236...97J}
\begin{barticle}
\bauthor{\bsnm{{Jing}}, \binits{J.}},
\bauthor{\bsnm{{Lee}}, \binits{J.}},
\bauthor{\bsnm{{Spirock}}, \binits{T.J.}},
\bauthor{\bsnm{{Wang}}, \binits{H.}}:
\byear{2006},
\batitle{{Periodic Motion Along Solar Filaments}}.
\bjtitle{\solphys}
\bvolume{236},
\bfpage{97}.
\doiurl{10.1007/s11207-006-0126-1}.
\adsurl{2006SoPh..236...97J}.
\end{barticle}
\endbibitem

\bibitem[\protect\citeauthoryear{{Kleczek} and
  {Kuperus}}{1969}]{1969SoPh....6...72K}
\begin{barticle}
\bauthor{\bsnm{{Kleczek}}, \binits{J.}},
\bauthor{\bsnm{{Kuperus}}, \binits{M.}}:
\byear{1969},
\batitle{{Oscillatory Phenomena in Quiescent Prominences}}.
\bjtitle{\solphys}
\bvolume{6},
\bfpage{72}.
\doiurl{10.1007/BF00146797}.
\adsurl{1969SoPh....6...72K}.
\end{barticle}
\endbibitem

\bibitem[\protect\citeauthoryear{{Labrosse}
  \textit{et~al.}}{2010}]{2010SSRv..151..243L}
\begin{barticle}
\bauthor{\bsnm{{Labrosse}}, \binits{N.}},
\bauthor{\bsnm{{Heinzel}}, \binits{P.}},
\bauthor{\bsnm{{Vial}}, \binits{J.-C.}},
\bauthor{\bsnm{{Kucera}}, \binits{T.}},
\bauthor{\bsnm{{Parenti}}, \binits{S.}},
\bauthor{\bsnm{{Gun{\'a}r}}, \binits{S.}},
\bauthor{\bsnm{{Schmieder}}, \binits{B.}},
\bauthor{\bsnm{{Kilper}}, \binits{G.}}:
\byear{2010},
\batitle{{Physics of Solar Prominences: I Spectral Diagnostics and
  Non-LTE Modelling}}.
\bjtitle{\ssr}
\bvolume{151},
\bfpage{243}.
\doiurl{10.1007/s11214-010-9630-6}.
\adsurl{2010SSRv..151..243L}.
\end{barticle}
\endbibitem

\bibitem[\protect\citeauthoryear{{Lemen}
  \textit{et~al.}}{2012}]{2012SoPh..275...17L}
\begin{barticle}
\bauthor{\bsnm{{Lemen}}, \binits{J.R.}},
\bauthor{\bsnm{{Title}}, \binits{A.M.}},
\bauthor{\bsnm{{Akin}}, \binits{D.J.}},
\bauthor{\bsnm{{Boerner}}, \binits{P.F.}},
\bauthor{\bsnm{{Chou}}, \binits{C.}},
\bauthor{\bsnm{{Drake}}, \binits{J.F.}},
\bauthor{\bsnm{{Duncan}}, \binits{D.W.}},
\bauthor{\bsnm{{Edwards}}, \binits{C.G.}},
\bauthor{\bsnm{{Friedlaender}}, \binits{F.M.}},
\bauthor{\bsnm{{Heyman}}, \binits{G.F.}},
\bauthor{\bsnm{{Hurlburt}}, \binits{N.E.}},
\bauthor{\bsnm{{Katz}}, \binits{N.L.}},
\bauthor{\bsnm{{Kushner}}, \binits{G.D.}},
\bauthor{\bsnm{{Levay}}, \binits{M.}},
\bauthor{\bsnm{{Lindgren}}, \binits{R.W.}},
\bauthor{\bsnm{{Mathur}}, \binits{D.P.}},
\bauthor{\bsnm{{McFeaters}}, \binits{E.L.}},
\bauthor{\bsnm{{Mitchell}}, \binits{S.}},
\bauthor{\bsnm{{Rehse}}, \binits{R.A.}},
\bauthor{\bsnm{{Schrijver}}, \binits{C.J.}},
\bauthor{\bsnm{{Springer}}, \binits{L.A.}},
\bauthor{\bsnm{{Stern}}, \binits{R.A.}},
\bauthor{\bsnm{{Tarbell}}, \binits{T.D.}},
\bauthor{\bsnm{{Wuelser}}, \binits{J.-P.}},
\bauthor{\bsnm{{Wolfson}}, \binits{C.J.}},
\bauthor{\bsnm{{Yanari}}, \binits{C.}},
\bauthor{\bsnm{{Bookbinder}}, \binits{J.A.}},
\bauthor{\bsnm{{Cheimets}}, \binits{P.N.}},
\bauthor{\bsnm{{Caldwell}}, \binits{D.}},
\bauthor{\bsnm{{Deluca}}, \binits{E.E.}},
\bauthor{\bsnm{{Gates}}, \binits{R.}},
\bauthor{\bsnm{{Golub}}, \binits{L.}},
\bauthor{\bsnm{{Park}}, \binits{S.}},
\bauthor{\bsnm{{Podgorski}}, \binits{W.A.}},
\bauthor{\bsnm{{Bush}}, \binits{R.I.}},
\bauthor{\bsnm{{Scherrer}}, \binits{P.H.}},
\bauthor{\bsnm{{Gummin}}, \binits{M.A.}},
\bauthor{\bsnm{{Smith}}, \binits{P.}},
\bauthor{\bsnm{{Auker}}, \binits{G.}},
\bauthor{\bsnm{{Jerram}}, \binits{P.}},
\bauthor{\bsnm{{Pool}}, \binits{P.}},
\bauthor{\bsnm{{Soufli}}, \binits{R.}},
\bauthor{\bsnm{{Windt}}, \binits{D.L.}},
\bauthor{\bsnm{{Beardsley}}, \binits{S.}},
\bauthor{\bsnm{{Clapp}}, \binits{M.}},
\bauthor{\bsnm{{Lang}}, \binits{J.}},
\bauthor{\bsnm{{Waltham}}, \binits{N.}}:
\byear{2012},
\batitle{{The Atmospheric Imaging Assembly (AIA) on the Solar Dynamics
  Observatory (SDO)}}.
\bjtitle{\solphys}
\bvolume{275},
\bfpage{17}.
\doiurl{10.1007/s11207-011-9776-8}.
\adsurl{2012SoPh..275...17L}.
\end{barticle}
\endbibitem

\bibitem[\protect\citeauthoryear{{Li} and {Zhang}}{2012}]{2012ApJ...760L..10L}
\begin{barticle}
\bauthor{\bsnm{{Li}}, \binits{T.}},
\bauthor{\bsnm{{Zhang}}, \binits{J.}}:
\byear{2012},
\batitle{{SDO/AIA Observations of Large-amplitude Longitudinal Oscillations in
  a Solar Filament}}.
\bjtitle{\apjl}
\bvolume{760},
\bfpage{L10}.
\doiurl{10.1088/2041-8205/760/1/L10}.
\adsurl{2012ApJ...760L..10L}.
\end{barticle}
\endbibitem

\bibitem[\protect\citeauthoryear{{Luna} and
  {Karpen}}{2012}]{2012ApJ...750L...1L}
\begin{barticle}
\bauthor{\bsnm{{Luna}}, \binits{M.}},
\bauthor{\bsnm{{Karpen}}, \binits{J.}}:
\byear{2012},
\batitle{{Large-amplitude Longitudinal Oscillations in a Solar Filament}}.
\bjtitle{\apjl}
\bvolume{750},
\bfpage{L1}.
\doiurl{10.1088/2041-8205/750/1/L1}.
\adsurl{2012ApJ...750L...1L}.
\end{barticle}
\endbibitem

\bibitem[\protect\citeauthoryear{{Luna}, {D{\'{\i}}az}, and
  {Karpen}}{2012}]{2012ApJ...757...98L}
\begin{barticle}
\bauthor{\bsnm{{Luna}}, \binits{M.}},
\bauthor{\bsnm{{D{\'{\i}}az}}, \binits{A.J.}},
\bauthor{\bsnm{{Karpen}}, \binits{J.}}:
\byear{2012},
\batitle{{The Effects of Magnetic-field Geometry on Longitudinal Oscillations
  of Solar Prominences}}.
\bjtitle{\apj}
\bvolume{757},
\bfpage{98}.
\doiurl{10.1088/0004-637X/757/1/98}.
\adsurl{2012ApJ...757...98L}.
\end{barticle}
\endbibitem

\bibitem[\protect\citeauthoryear{{Luna}
  \textit{et~al.}}{2014}]{2014ApJ...785...79L}
\begin{barticle}
\bauthor{\bsnm{{Luna}}, \binits{M.}},
\bauthor{\bsnm{{Knizhnik}}, \binits{K.}},
\bauthor{\bsnm{{Muglach}}, \binits{K.}},
\bauthor{\bsnm{{Karpen}}, \binits{J.}},
\bauthor{\bsnm{{Gilbert}}, \binits{H.}},
\bauthor{\bsnm{{Kucera}}, \binits{T.A.}},
\bauthor{\bsnm{{Uritsky}}, \binits{V.}}:
\byear{2014},
\batitle{{Observations and Implications of Large-amplitude Longitudinal
  Oscillations in a Solar Filament}}.
\bjtitle{\apj}
\bvolume{785},
\bfpage{79}.
\doiurl{10.1088/0004-637X/785/1/79}.
\adsurl{2014ApJ...785...79L}.
\end{barticle}
\endbibitem

\bibitem[\protect\citeauthoryear{{Luna}
  \textit{et~al.}}{2016}]{2016ApJ...817..157L}
\begin{barticle}
\bauthor{\bsnm{{Luna}}, \binits{M.}},
\bauthor{\bsnm{{Terradas}}, \binits{J.}},
\bauthor{\bsnm{{Khomenko}}, \binits{E.}},
\bauthor{\bsnm{{Collados}}, \binits{M.}},
\bauthor{\bsnm{{de Vicente}}, \binits{A.}}:
\byear{2016},
\batitle{{On the Robustness of the Pendulum Model for Large-amplitude
  Longitudinal Oscillations in Prominences}}.
\bjtitle{\apj}
\bvolume{817},
\bfpage{157}.
\doiurl{10.3847/0004-637X/817/2/157}.
\adsurl{2016ApJ...817..157L}.
\end{barticle}
\endbibitem

\bibitem[\protect\citeauthoryear{{Mackay}
  \textit{et~al.}}{2010}]{2010SSRv..151..333M}
\begin{barticle}
\bauthor{\bsnm{{Mackay}}, \binits{D.H.}},
\bauthor{\bsnm{{Karpen}}, \binits{J.T.}},
\bauthor{\bsnm{{Ballester}}, \binits{J.L.}},
\bauthor{\bsnm{{Schmieder}}, \binits{B.}},
\bauthor{\bsnm{{Aulanier}}, \binits{G.}}:
\byear{2010},
\batitle{{Physics of Solar Prominences: II Magnetic Structure and
  Dynamics}}.
\bjtitle{\ssr}
\bvolume{151},
\bfpage{333}.
\doiurl{10.1007/s11214-010-9628-0}.
\adsurl{2010SSRv..151..333M}.
\end{barticle}
\endbibitem

\bibitem[\protect\citeauthoryear{{Markwardt}}{2009}]{2009ASPC..411..251M}
\begin{bchapter}
\bauthor{\bsnm{{Markwardt}}, \binits{C.B.}}:
\byear{2009},
\bctitle{{Non-linear Least-squares Fitting in IDL with MPFIT}}.
In: \beditor{\bsnm{{Bohlender}}, \binits{D.A.}},
\beditor{\bsnm{{Durand}}, \binits{D.}},
\beditor{\bsnm{{Dowler}}, \binits{P.}} (eds.)
\bbtitle{Astronomical Data Analysis Software and Systems XVIII},
\bsertitle{Astronomical Society of the Pacific Conference Series}
\bseriesno{411},
\bfpage{251}.
\adsurl{2009ASPC..411..251M}.
\end{bchapter}
\endbibitem

\bibitem[\protect\citeauthoryear{{Nakariakov} and
  {Verwichte}}{2005}]{2005LRSP....2....3N}
\begin{barticle}
\bauthor{\bsnm{{Nakariakov}}, \binits{V.M.}},
\bauthor{\bsnm{{Verwichte}}, \binits{E.}}:
\byear{2005},
\batitle{{Coronal Waves and Oscillations}}.
\bjtitle{Living Reviews in Solar Physics}
\bvolume{2}.
\doiurl{10.12942/lrsp-2005-3}.
\adsurl{2005LRSP....2....3N}.
\end{barticle}
\endbibitem

\bibitem[\protect\citeauthoryear{{Okamoto}
  \textit{et~al.}}{2004}]{2004ApJ...608.1124O}
\begin{barticle}
\bauthor{\bsnm{{Okamoto}}, \binits{T.J.}},
\bauthor{\bsnm{{Nakai}}, \binits{H.}},
\bauthor{\bsnm{{Keiyama}}, \binits{A.}},
\bauthor{\bsnm{{Narukage}}, \binits{N.}},
\bauthor{\bsnm{{UeNo}}, \binits{S.}},
\bauthor{\bsnm{{Kitai}}, \binits{R.}},
\bauthor{\bsnm{{Kurokawa}}, \binits{H.}},
\bauthor{\bsnm{{Shibata}}, \binits{K.}}:
\byear{2004},
\batitle{{Filament Oscillations and Moreton Waves Associated with EIT Waves}}.
\bjtitle{\apj}
\bvolume{608},
\bfpage{1124}.
\doiurl{10.1086/420838}.
\adsurl{2004ApJ...608.1124O}.
\end{barticle}
\endbibitem

\bibitem[\protect\citeauthoryear{{Oliver} and
  {Ballester}}{2002}]{2002SoPh..206...45O}
\begin{barticle}
\bauthor{\bsnm{{Oliver}}, \binits{R.}},
\bauthor{\bsnm{{Ballester}}, \binits{J.L.}}:
\byear{2002},
\batitle{{Oscillations in Quiescent Solar Prominences Observations and Theory
  (Invited Review)}}.
\bjtitle{\solphys}
\bvolume{206},
\bfpage{45}.
\doiurl{10.1023/A:1014915428440}.
\adsurl{2002SoPh..206...45O}.
\end{barticle}
\endbibitem

\bibitem[\protect\citeauthoryear{{Pant}
  \textit{et~al.}}{2015}]{2015RAA....15.1713P}
\begin{barticle}
\bauthor{\bsnm{{Pant}}, \binits{V.}},
\bauthor{\bsnm{{Srivastava}}, \binits{A.K.}},
\bauthor{\bsnm{{Banerjee}}, \binits{D.}},
\bauthor{\bsnm{{Goossens}}, \binits{M.}},
\bauthor{\bsnm{{Chen}}, \binits{P.-F.}},
\bauthor{\bsnm{{Joshi}}, \binits{N.C.}},
\bauthor{\bsnm{{Zhou}}, \binits{Y.-H.}}:
\byear{2015},
\batitle{{MHD Seismology of a loop-like filament tube by observed kink waves}}.
\bjtitle{Research in Astronomy and Astrophysics}
\bvolume{15},
\bfpage{1713}.
\doiurl{10.1088/1674-4527/15/10/008}.
\adsurl{2015RAA....15.1713P}.
\end{barticle}
\endbibitem

\bibitem[\protect\citeauthoryear{{Ramsey} and
  {Smith}}{1966}]{1966AJ.....71..197R}
\begin{barticle}
\bauthor{\bsnm{{Ramsey}}, \binits{H.E.}},
\bauthor{\bsnm{{Smith}}, \binits{S.F.}}:
\byear{1966},
\batitle{{Flare-initiated filamei it oscillations}}.
\bjtitle{\aj}
\bvolume{71},
\bfpage{197}.
\doiurl{10.1086/109903}.
\adsurl{1966AJ.....71..197R}.
\end{barticle}
\endbibitem

\bibitem[\protect\citeauthoryear{{Ruderman} and
  {Luna}}{2016}]{2016arXiv160503376R}
\begin{botherref}
\oauthor{\bsnm{{Ruderman}}, \binits{M.}},
\oauthor{\bsnm{{Luna}}, \binits{M.}}:
2016,
{Damping of prominence longitudinal oscillations due to mass accretion}.
\textit{ArXiv e-prints}.
\adsurl{2016arXiv160503376R}.
\end{botherref}
\endbibitem

\bibitem[\protect\citeauthoryear{{Ruderman} and
  {Erd{\'e}lyi}}{2009}]{2009SSRv..149..199R}
\begin{barticle}
\bauthor{\bsnm{{Ruderman}}, \binits{M.S.}},
\bauthor{\bsnm{{Erd{\'e}lyi}}, \binits{R.}}:
\byear{2009},
\batitle{{Transverse Oscillations of Coronal Loops}}.
\bjtitle{\ssr}
\bvolume{149},
\bfpage{199}.
\doiurl{10.1007/s11214-009-9535-4}.
\adsurl{2009SSRv..149..199R}.
\end{barticle}
\endbibitem

\bibitem[\protect\citeauthoryear{{Schmieder}
  \textit{et~al.}}{2013}]{2013ApJ...777..108S}
\begin{barticle}
\bauthor{\bsnm{{Schmieder}}, \binits{B.}},
\bauthor{\bsnm{{Kucera}}, \binits{T.A.}},
\bauthor{\bsnm{{Knizhnik}}, \binits{K.}},
\bauthor{\bsnm{{Luna}}, \binits{M.}},
\bauthor{\bsnm{{Lopez-Ariste}}, \binits{A.}},
\bauthor{\bsnm{{Toot}}, \binits{D.}}:
\byear{2013},
\batitle{{Propagating Waves Transverse to the Magnetic Field in a Solar
  Prominence}}.
\bjtitle{\apj}
\bvolume{777},
\bfpage{108}.
\doiurl{10.1088/0004-637X/777/2/108}.
\adsurl{2013ApJ...777..108S}.
\end{barticle}
\endbibitem

\bibitem[\protect\citeauthoryear{{Shen}
  \textit{et~al.}}{2014}]{2014ApJ...795..130S}
\begin{barticle}
\bauthor{\bsnm{{Shen}}, \binits{Y.}},
\bauthor{\bsnm{{Liu}}, \binits{Y.D.}},
\bauthor{\bsnm{{Chen}}, \binits{P.F.}},
\bauthor{\bsnm{{Ichimoto}}, \binits{K.}}:
\byear{2014},
\batitle{{Simultaneous Transverse Oscillations of a Prominence and a Filament
  and Longitudinal Oscillation of Another Filament Induced by a Single Shock
  Wave}}.
\bjtitle{\apj}
\bvolume{795},
\bfpage{130}.
\doiurl{10.1088/0004-637X/795/2/130}.
\adsurl{2014ApJ...795..130S}.
\end{barticle}
\endbibitem

\bibitem[\protect\citeauthoryear{{Tripathi}, {Isobe}, and
  {Jain}}{2009}]{2009SSRv..149..283T}
\begin{barticle}
\bauthor{\bsnm{{Tripathi}}, \binits{D.}},
\bauthor{\bsnm{{Isobe}}, \binits{H.}},
\bauthor{\bsnm{{Jain}}, \binits{R.}}:
\byear{2009},
\batitle{{Large Amplitude Oscillations in Prominences}}.
\bjtitle{\ssr}
\bvolume{149},
\bfpage{283}.
\doiurl{10.1007/s11214-009-9583-9}.
\adsurl{2009SSRv..149..283T}.
\end{barticle}
\endbibitem

\bibitem[\protect\citeauthoryear{{Vr{\v s}nak}
  \textit{et~al.}}{2007}]{2007A&A...471..295V}
\begin{barticle}
\bauthor{\bsnm{{Vr{\v s}nak}}, \binits{B.}},
\bauthor{\bsnm{{Veronig}}, \binits{A.M.}},
\bauthor{\bsnm{{Thalmann}}, \binits{J.K.}},
\bauthor{\bsnm{{{\v Z}ic}}, \binits{T.}}:
\byear{2007},
\batitle{{Large amplitude oscillatory motion along a solar filament}}.
\bjtitle{\aap}
\bvolume{471},
\bfpage{295}.
\doiurl{10.1051/0004-6361:20077668}.
\adsurl{2007A\%26A...471..295V}.
\end{barticle}
\endbibitem

\bibitem[\protect\citeauthoryear{{Yuan} and {Van
  Doorsselaere}}{2016}]{2016ApJS..223...23Y}
\begin{barticle}
\bauthor{\bsnm{{Yuan}}, \binits{D.}},
\bauthor{\bsnm{{Van Doorsselaere}}, \binits{T.}}:
\byear{2016},
\batitle{{Forward Modeling of Standing Kink Modes in Coronal Loops. I.
  Synthetic Views}}.
\bjtitle{\apjs}
\bvolume{223},
\bfpage{23}.
\doiurl{10.3847/0067-0049/223/2/23}.
\adsurl{2016ApJS..223...23Y}.
\end{barticle}
\endbibitem

\bibitem[\protect\citeauthoryear{{Zhang}
  \textit{et~al.}}{2012}]{2012A&A...542A..52Z}
\begin{barticle}
\bauthor{\bsnm{{Zhang}}, \binits{Q.M.}},
\bauthor{\bsnm{{Chen}}, \binits{P.F.}},
\bauthor{\bsnm{{Xia}}, \binits{C.}},
\bauthor{\bsnm{{Keppens}}, \binits{R.}}:
\byear{2012},
\batitle{{Observations and simulations of longitudinal oscillations of an
  active region prominence}}.
\bjtitle{\aap}
\bvolume{542},
\bfpage{A52}.
\doiurl{10.1051/0004-6361/201218786}.
\adsurl{2012A\%26A...542A..52Z}.
\end{barticle}
\endbibitem

\end{thebibliography}

\end{article} 

\end{document}